\documentclass[12pt]{iopart}

\usepackage{amssymb}
\usepackage{graphicx}

\begin{document}

\title{Phonons in graphene with point defects}

\author{Vadym Adamyan and Vladimir Zavalniuk}
\address{Department of Theoretical Physics, Odessa I I Mechnikov National University,\\2 Dvoryanskaya St., Odessa 65026, Ukraine}
\eads{\mailto{vadamyan@onu.edu.ua}, \mailto{vzavalnyuk@onu.edu.ua}}

\begin{abstract}
The phonon density of states (DOS) of graphene with different types of point defects (carbon isotopes, substitution atoms, vacancies) is considered.
Using a solvable model which is based on the harmonic approximation and the assumption that the elastic forces act only between nearest neighboring ions we calculate corrections to graphene DOS dependent on type and concentration of defects. In particular the correction due to isotopic dimers is determined. It is shown that a relatively small concentration of defects may lead to significant and specific changes in the DOS, especially at low frequencies, near the Van Hove points and in the vicinity of the K-points of the Brillouin zone. In some cases defects generate one or several narrow gaps near the critical points of the phonon DOS as well as resonance states in the Brillouin zone regular points. All types of defects are characterized by the appearance of one or more additional Van Hove peaks near the (Dirac) K points and their singular contribution may be comparable with the effect of electron-phonon interaction. Besides, for low frequencies and near the critical points the relative change in density of states may be many times higher than the concentration of defects.
\end{abstract}

\submitto{\JPCM}

\maketitle
% ----------------------------------------------------------------
\section{Introduction}

Due to the low atomic mass of carbon and high binding energy of the valent $sp_2$ bonds \cite{CoulsonValence}, graphene-based nanostructures (single- or several-layered graphene and nanotubes of different kinds) have high rigidity in one direction combined with excellent flexibility in the others \cite{Demczyk,Salvetat} leading to the extraordinary sound velocity (about 20000 km/sec) \cite{Maultzsch,TA1,TA2} and thermal conductivity \cite{HoneWhitney,SaitoDresselhaus,UnderstandingCNTs,SaitoZettl}. Due to these properties carbon compounds can be used not only for plastic and hydrocarbon resin reinforcement, but also as one of the basis materials for nano-mechanics and nano-electronics \cite{CumingsZettl,TuXu}.

It is natural to expect that as well as in bulk crystals even small concentrations of point defects  in graphene-based $1D$ and $2D$ nanostructures (directly observed in \cite{DefectsDir1,DefectsDir2}) may lead to specific shifts, broadenings and additional characteristic singularities in the electron and phonon densities of states and thus change their optical absorption, low-temperature specific heat and transport properties. Such effects in graphene and carbon nanotubes are significant because of the occurrence of three isotopes in natural carbon ($\,^{12}C,\,^{13}C,\,^{14}C$) with the part of the $\,^{13}C$ isotope in the chemically pure carbon exceeding one per cent and also because of the high solubility of substitution defects of trivalent atoms (such as aluminium, boron and nitrogen) and monovalent atom adsorption susceptibility. According to \cite{FreeMonolayers}, the defect density in a graphene monolayer stabilized on a substrate can reach several per cent and for some applications it can be additionally doped to raise the electrical conductivity. It is shown that unintentional doping of pristine unprocessed graphene under ambient conditions can reach as high as 1\% \cite{DefectsAmb}, while the highest achieved doping level of N is about 5\% \cite{DefectsNitrogen5p}. It was also computationally established that the graphene planar structure is kept even for 12\% Al and 20\% N concentrations \cite{DefectsAL12p,DefectsNitrogen20p}. In addition, most of the chemically adsorbed atoms (especially monovalent ones) can be treated as isotopic defects because they are bound to carbon atoms by the $\pi$-electron bonds which are not involved in lattice formation (this may need slight correction of $\sigma$ electrons' binding energies and angles, but as the first approximation they can be taken as in the ideal graphene). It is worth mentioning that the impact of defects on the electronic properties of graphite, graphene and carbon nanotubes along with a detailed investigation of how electron-phonon interaction affects the phonon dispersion curves especially near the K points of the Brillouin zone were thoroughly studied in a large number of works \cite{SaitoDresselhaus,UnderstandingCNTs,SaitoZettl,CastroNeto,Charlier,RocheJiangFoaTorres,MaultzschEF,PiscanecEF}, while too little attention has been paid so far to description of the direct influence of defects on the phonon spectra. However the anomalies of the phonon spectra due to the electron-phonon interaction may be visibly distorted by the Van Hove spikes induced by defects.

This paper is devoted to the description of the effect of some point defects (isotopic defects, substitutional atoms, vacant lattice sites) on the phonon spectrum of graphene. As a starting point we consider in Section 2 the ideal graphene phonon spectrum in the simple harmonic approximation assuming that elastic forces act only between nearest neighboring hard ions and are described by three harmonic force constants $J_{1},J_{2},J_{3}$ corresponding to three different parts of the interatomic interaction: the central(1) and non-central(2) in-plane forces and the empirical to-the-plane backmoving non-central force(3). The values of these constants are chosen to get the least discrepancy of the eigenfrequencies calculated in the framework of a simple model for the $\Gamma$, K and M-points of the Brillouin zone  with those obtained in well-performed inelastic x-ray scattering experiments \cite{Maultzsch}. It is important to underscore that a convergence within the limits of experimental error of theoretical and experimental phonon dispersion curves for graphene is unlikely to be attainable on the base of the simple three-parameter nearest-neighbors model. For example, by using of such a model it is impossible to explain the phenomenon of "overbending", which was observed on the dispersion curves for graphite and graphene \cite{Maultzsch,MaultzschEF,PiscanecEF,HREELS}. However, being operationally rather simple, it gives for appropriate values of the constants $J_{1},J_{2},J_{3}$ the dispersion curves and phonon density of states (DOS) for graphene, which are qualitatively just similar and quantitatively are rather close to those obtained using theoretical models with greater number of force constants \cite{Maultzsch, Falkovsky,ltp1}. On the base of the three-parameter simplified model for the ideal graphene we study further the effect of point defects on the graphene phonon spectra.

In Section 3 we describe the contribution of isotopic defects to the graphene phonon DOS, analyzing separately the cases of single defect, dimer and pair of distant defects. In doing so we specify for graphene approaches based on the method of classic Green functions which were developed more than fifty years ago in \cite{Lifshits,LifshitsStepanova,Elliott,Maradudin}.

In Section 4 the same problem is considered for other point defects: substitutional atoms and vacancies. In these more complicated cases along with masses of defects the force constants between defect sites and their nearest neighbors should be changed.

Finally in section 5 we discuss characteristic traces of considered point defects in optical spectra and the heat capacity of graphene.

As illustrations we demonstrate linear in defect concentration contributions to DOS as more important for comparison with real experiments and estimations of defects manifestation in the phonon spectra.

\section{Phonons in ideal graphene}

\begin{figure}[!hbp]
\center
\includegraphics[scale=0.75]{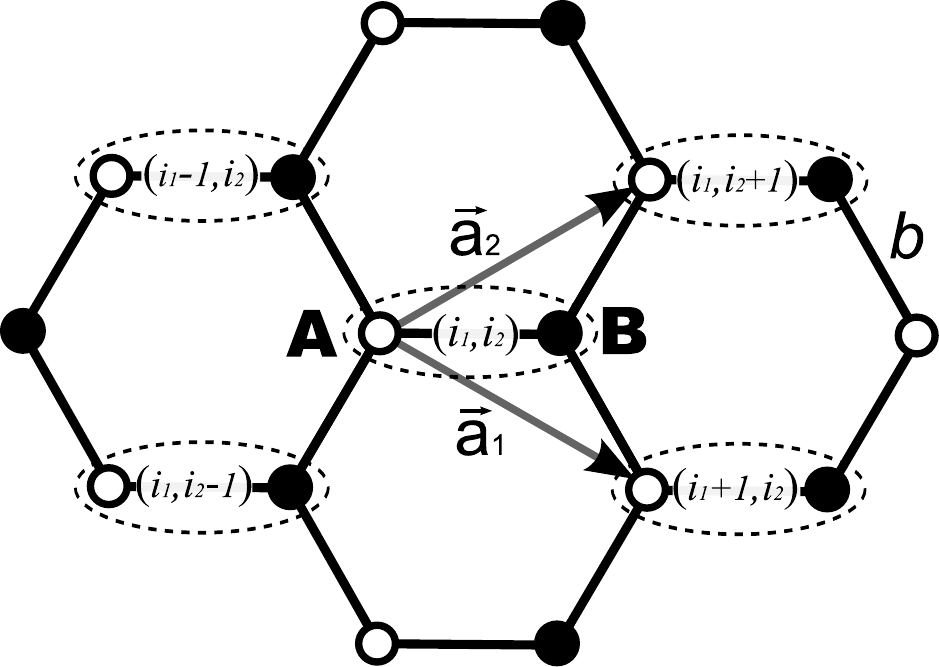}
\caption{The graphitic plane structure. $\mathbf{A}$ (full circles) and $\mathbf{B}$ (hollow circles) represent two sublattices, $\mathbf{a}_1$ and $\mathbf{a}_2$ are two primitive translation vectors, $|\mathbf{a}_{1,2}|=a=\sqrt{3} b$, where $b=0.142nm$ is the interatomic distance or the length of carbon-carbon $\sigma$-bond.}\label{picCell}
\end{figure}

The ideal graphene is a 2D crystal with two carbon atoms per elementary cell (further we will call them "$A$" and "$B$") (figure \ref{picCell}).

The equilibrium position of atoms on sublattices of the carbon plane can described by vectors
\[
\mathbf{R}_{\mathbf{n},A}^{0}=n_{1,A}\mathbf{a}_{1}+ n_{2,A}\mathbf{a}_{2}, \qquad
\mathbf{R}_{\mathbf{n},B}^{0}=n_{1,B}\mathbf{a}_{1}+ n_{2,B}\mathbf{a}_{2}+\frac{1}{\sqrt{3}}\left(\mathbf{a}_{1}+\mathbf{a}_{2}\right)
 \]
with integer $n_{1,A},n_{2,A};n_{1,B},n_{2,B}$. The instantaneous ion configuration of the graphene plane is characterized by the actual positions of atoms
\[  \mathbf{R}_{\mathbf{n}\sigma}=\mathbf{R}_{\mathbf{n}\sigma}^{0}+\mathbf{u}_{\mathbf{n}\sigma}(t), \quad \sigma=(A,B), \]
with time-dependent displacements
\[
\mathbf{u}_{\mathbf{n},\sigma}(t)=u_{\mathbf{n}\sigma,1}(t)\mathbf{a}_{1}+ u_{\mathbf{n}\sigma,2}(t)\mathbf{a}_{2}+u_{\mathbf{n}\sigma,3}(t)\mathbf{a}_{3}
\]
around $\mathbf{R}_{\mathbf{n}\sigma}^{0}$, where $\mathbf{a}_{3}$ is the orthogonal to the lattice plane unit vector. Since the angle between translation vectors $\mathbf{a}_{1}$ and $\mathbf{a}_{2}$ is $\pi/3$, then the scalar product of displacement vectors $\mathbf{u}_{\mathbf{n}\sigma}$ and $\mathbf{u}_{\mathbf{n}'\sigma'}$ has the form
\[
\begin{array}{l} \left(\mathbf{u}_{\mathbf{n}\sigma},\mathbf{u}_{\mathbf{n}'\sigma'}\right)= \\
\quad u_{\mathbf{n}\sigma,1}u_{\mathbf{n}'\sigma',1}
+\frac{1}{2}\left(u_{\mathbf{n}\sigma,1}u_{\mathbf{n}'\sigma',2}+u_{\mathbf{n}\sigma,2}u_{\mathbf{n}'\sigma',1}\right)
+u_{\mathbf{n}\sigma,2}u_{\mathbf{n}'\sigma',2}+u_{\mathbf{n}\sigma,3}u_{\mathbf{n}'\sigma',3}.
\end{array}
\]
Therefore the kinetic energy $K$ of lattice atoms can be written as follows
\[\begin{array}{l} K=\frac{1}{2}\sum\limits_{\mathbf{n}}\sum\limits_{\sigma}^{}m_{0}\mathbf{\dot{u}}_{\mathbf{n},\sigma}^{2} = \\
\qquad \frac{1}{2}\sum\limits_{\mathbf{n}}\sum\limits_{\sigma}^{}m_{0} \left(\dot{u}_{\mathbf{n}\sigma,1}^{2}+\dot{u}_{\mathbf{n}\sigma,2}^{2}+\dot{u}_{\mathbf{n}\sigma,1}\dot{u}_{\mathbf{n}\sigma,2}
+\dot{u}_{\mathbf{n}\sigma,3}^2 \right), \end{array} \]
where $m_{0}$ is the mass of the carbon atom (isotope $\,^{12}C$).

The total potential energy $W$ of ideal graphene is modeled as a sum of three components determined by three different force constants:
\begin{equation}\label{pot}
W=W_{c}+W_{i.p.nc}+W_{o.p.nc},
\end{equation}
where $W_{c}$ is the part of total potential energy depending on the relative displacements of neighboring atoms along the lines connecting their equilibrium positions i.e. the part of the potential energy due to the central forces, $W_{i.p. nc}$ is the component depending only on magnitudes of in-plane relative displacements of interacting nearest neighbors,  $W_{o.p. nc}$ is the non-central component conditioned by  $\pi$-electrons interaction, which depends on the out-of-plane relative displacements of neighboring atoms.  For our parametrization the nearest neighbors to a site $\mathbf{n}A$ $(\mathbf{n}B)$ of the graphene lattice are sites $(\mathbf{n}+\mathbf{S})B$ $((\mathbf{n}-\mathbf{S})A)$, where $\mathbf{S}$ either $0$-vector, or $-\mathbf{a}_{1}$, or $-\mathbf{a}_{2}$.
Using this notation and setting
\[
\Delta\mathbf{u}_{\mathbf{nS},BA}=\mathbf{u}_{\mathbf{n}+\mathbf{S},B}-\mathbf{u}_{\mathbf{n},A}, \qquad
\Delta\mathbf{u}_{\mathbf{nS},AB}=\mathbf{u}_{\mathbf{n}-\mathbf{S},A}-\mathbf{u}_{\mathbf{n},B}
\]
we get the following expressions for the components of potential energy in (\ref{pot})
\begin{equation}\label{pot1}
\begin{array}{l}
W_{c}\,=\, \frac{1}{4b^2}J_{1}\sum\limits_{\mathbf{n}}^{}\sum\limits_{\mathbf{S}}^{}\left[ \left(\Delta\mathbf{u}_{\mathbf{nS},BA},\mathbf{R}_{\mathbf{n}+\mathbf{S},B}^{0}-\mathbf{R}_{\mathbf{n},A}^{0}\right)^2 \right. + \\
\qquad\;\;+\left.\left(\Delta\mathbf{u}_{\mathbf{nS},AB},\mathbf{R}_{\mathbf{n}-\mathbf{S},A}^{0}-\mathbf{R}_{\mathbf{n},B}^{0}\right)^2 \right], \\
W_{i.p.nc}=\frac{1}{4}J_2\sum\limits_{\mathbf{n}}^{}\sum\limits_{\mathbf{S}}^{}\left[\left[\Delta\mathbf{u}_{\mathbf{nS},BA} \times \mathbf{a}_3\right]^2 + \left[\Delta\mathbf{u}_{\mathbf{nS},AB} \times \mathbf{a}_3\right]^2 \right], \\
W_{o.p.nc}=\frac{1}{4}J_3\sum\limits_{\mathbf{n}}^{}\sum\limits_{\mathbf{S}}^{}
\left[\left(\Delta\mathbf{u}_{\mathbf{nS},BA}\cdot\mathbf{a}_3\right)^2 + \left(\Delta\mathbf{u}_{\mathbf{nS},AB}\cdot \mathbf{a}_3\right)^2 \right].
\end{array}
\end{equation}
with indeterminate force constants $J_{1},J_{2},J_{3}$.

As usual in lattice dynamics, we use further the Bloch theorem, according to which the atom or ion displacements on sublattice sites $\mathbf{n}$ for the normal modes differ only by phase factors $\exp{\left[\rmi\mathbf{k}\cdot\mathbf{R}_{\mathbf{n}}^{0}\right]}$, where $\mathbf{k}=(k_{1},k_{2})$ run the Brillouin zone of the reciprocal sublattice. In this way we obtain that the squares of the frequencies $\omega_{j}(\mathbf{k})$ for the different branches $j$ of the ideal graphene phonon spectra coincide with eigenvalues of the \textit{dynamical matrix}
$$ \mathbf{D}(\mathbf{k})=\mathbf{M}^{-1/2}\left(J_{1}\mathbf{G}_{c}+J_{2}\mathbf{G}_{i.p.nc}+J_{3}\mathbf{G}_{o.p.nc}\right)
\mathbf{M}^{-1/2},
$$
where  $\mathbf{M}$ is the mass matrix:
\[
\mathbf{M}=\left(\begin{array}{ccc}
                   M_{0} & 0 \\
                   0     & M_{0}
                 \end{array}
           \right),
\qquad\quad
\mathbf{M}_{0}=m_{0}(\mathbf{M}_{i.p.}+\mathbf{M}_{o.p.}),
\]
\[
\mathbf{M}_{i.p.} = \left(\begin{array}{ccc}
                     1   & \frac{1}{2} & 0 \\
                     \frac{1}{2} &  1  & 0 \\
                     0   &  0  & 0
                    \end{array}\right),
\qquad
\mathbf{M}_{o.p.} = \left(\begin{array}{ccc}
                     0 & 0 & 0 \\
                     0 & 0 & 0 \\
                     0 & 0 & 1
                    \end{array}\right),
\]
and $\mathbf{G}_{c}$,$\mathbf{G}_{i.p.nc}$ and $\mathbf{G}_{o.p.nc}$ are Hermitian $2\times$-block matrices of the form
\[
\mathbf{G}=\left(\begin{array}{cc}
                     \mathbf{G}^{d}   & \mathbf{G}^{a}  \\
                     \overline{\mathbf{G}^{a}}   &  \mathbf{G}^{d}
                   \end{array}
             \right)
\]
with $3 \times3$ diagonal and anti-diagonal blocks $\mathbf{G}^{d}$ and $\mathbf{G}^{a}$ , respectively:
\[
\begin{array}{ll}
\mathbf{G}_{c}^{d}=\frac{3}{2}\mathbf{M}_{i.p.},
&
\mathbf{G}_{c}^{a}=
-\frac{3}{2}\mathbf{M}_{i.p.} - \frac{3}{4}
        \left(\begin{array}{ccc}
         1-\rme^{\rmi k_{1}a} & 0 & 0  \\
         0 & 1-\rme^{\rmi k_{2}a} & 0  \\
         0 & 0 & 0
        \end{array}
   \right),
\\
\mathbf{G}_{i.p.nc}^{d} = 3 \mathbf{M}_{i.p.},
&
\mathbf{G}_{i.p.nc}^{a}=
-\left(1+\rme^{\rmi k_{1}a}+\rme^{\rmi k_{2}a}\right) \mathbf{M}_{i.p.},
\\
\mathbf{G}_{o.p.nc}^{d}=3 \mathbf{M}_{o.p.},
&
\mathbf{G}_{o.p.nc}^{a}=
- \left(1+\rme^{\rmi k_{1}a}+\rme^{\rmi k_{2}a}\right) \mathbf{M}_{o.p.}.
\end{array}
\]

Hence for the ideal graphene plane there are six branches $\omega_{j}(\mathbf{k})$ ($j=LA,TA,ZA,LO,TO,ZO$) of phonon spectra: for two of them ($\omega_{ZA}(\mathbf{k})$ and $\omega_{ZO}(\mathbf{k})$) the atom displacements are perpendicular to the lattice plane while for the other four branches the atoms do not come out of the plane.

Setting
\begin{equation}\label{auxil}
\begin{array}{l}
F_0(\mathbf{k}) = 2 \left[\cos{(k_1 a-k_2 a)}+\cos{k_2 a}+\cos{k_1 a}\right], \\
F_1(\mathbf{k}) =  12 (J_1^2 + 2 J_1 J_2 + 2 J_2^2) +  F_0(\mathbf{k}) (J_1^2 + 8 J_1 J_2 + 8 J_2^2), \\
X_1 = J_1^2 + 16 J_1 J_2 + 16 J_2^2,  \qquad\quad  X_2 = J_1^2 - 8 J_1 J_2 - 8 J_2^2, \\
F_2(\mathbf{k}) =\left\{18 J_1^2 + 2 X_1 + 4 X_2 \cos{(k_1 a-k_2 a )}\left(\cos{k_1 a} + \cos{k_2 a}\right) \right. + \\
  \qquad\qquad \left. +  4 \cos{k_1 a} \cos{k_2 a} \left[X_2 + X_1 \cos{(k_1 a- k_2 a)}\right] - 6 J_1^2 F_0(\mathbf{k}) \right\}^\frac{1}{2},
\end{array}
\end{equation}
we get the following expressions for the eigenfrequencies $\omega_{j}(\mathbf{k})$:
\begin{equation}\label{eigen}
\begin{array}{l}
\omega_{ZA,ZO}(\mathbf{k}) = \left[ \frac{J_3}{m} \left(3 \pm \sqrt{3 + F_0(\mathbf{k})}\right) \right]^{1/2},\\
\omega_{LA,TA,LO,TO}(\mathbf{k}) = \left[ \frac{3 (J_1 + 2 J_2)}{2 m} \pm \frac{\sqrt{2}}{4 m} \sqrt{F_1(\mathbf{k}) \pm \sqrt{2}J_1 F_2(\mathbf{k})} \right]^{1/2}.
\end{array}
\end{equation}
The last expressions take a very simple form at the high symmetry points $\Gamma\,(\mathbf{k}a=(0,0))$, K $(\mathbf{k}a=(\frac{4}{3}\pi,\frac{2}{3}\pi))$ and M $(\mathbf{k}a=(\pi,\pi))$ of the graphene first Brillouin zone (\Tref{Table1}).

\Table{\label{Table1} Exact expressions for phonon frequencies (in $cm^{-1}$) in the $\Gamma$,M and K points of the Brillouin zone.}
\br
\; & \centre{1}{$\omega_1$}& \centre{1}{$\omega_2$}& \centre{1}{$\omega_3$}& \centre{1}{$\omega_4$}& \centre{1}{$\omega_5$}& \centre{1}{$\omega_6$}\\
\mr
$\Gamma$ & 0 & 0 & 0 & \centre{1}{$\sqrt{\frac{6 J_3}{m}}$} & \centre{2}{$\sqrt{\frac{3 (J_1 + 2 J_2)}{m}}$} \\
M        & $\sqrt{\frac{2 J_3}{m}}$ & $\sqrt{\frac{4 J_3}{m}}$ & $\sqrt{\frac{2 J_2}{m}}$ & $\sqrt{\frac{2(J_1 +    J_2)}{m}}$ & $\sqrt{\frac{J_1 + 4 J_2}{m}}$ & $\sqrt{\frac{3 J_1 + 4 J_2}{m}}$ \\
K        & \centre{2}{$\sqrt{\frac{3 J_3}{m}}$} & $\sqrt{\frac{3 J_2}{m}}$ & \centre{2}{$\sqrt{\frac{3(J_1 + 2 J_2)}{2m}}$} & $\sqrt{\frac{3(J_1 + J_2)}{m}}$ \\
\br
\endTable

In the vicinity of $\Gamma$-point there are three acoustic branches of the graphene phonon spectra, the eigenfrequencies of which in the limit $k\rightarrow 0$ do not depend on the direction of propagation. They can be distinguished by polarization of oscillations as the longitudinal branch $\omega_{LA}=c_{LA}k+O(k^{2})$ with longitudinal sound velocity
\begin{equation}\label{aclong}
c_{LA}= a\sqrt{ \frac{(J_1 + J_2)(J_1 + 4 J_2)}{6 (J_1+2 J_2) m_0} };
\end{equation}
the transverse in-plane branch $\omega_{TA}=c_{TA}k+O(k^{2})$ with in-plane transverse sound velocity
\begin{equation}\label{actin}
 c_{TA}=a\sqrt{ \frac{3 J_1 J_2 + 4 J_2^2}{6 (J_1+2 J_2) m_0} };
\end{equation}
the transverse out-of-plane branch $\omega_{ZA}=c_{ZA}k +O(k^{2})$ with transverse out-of-plane sound velocity
\begin{equation}\label{ac3}
c_{ZA}=a\sqrt{ \frac{J_3}{3 m_0} }.
\end{equation}

By definition the density of states $\rho (\omega)$ of any oscillatory system is given by expressions
\begin{equation}\label{first}
\begin{array}{l}
  \rho (\omega)=2\omega \mathcal{R}(\omega^{2}), \quad \omega >0,  \\
  \mathcal{R}(\lambda)=\sum\limits_{\nu}^{} \delta\left(\lambda - \omega_{\nu}^{2}\right) = \frac{1}{\pi}\lim\limits_{\varepsilon\downarrow 0} \sum\limits_{\nu}^{}\frac{\varepsilon}{\left(\lambda -\omega_{\nu}^{2}\right)^{2}+\varepsilon^{2}},
\end{array}
\end{equation}
where $\omega_{\nu}$ are eigenfrequencies of the system counted with respect to their multiplicities. In our case the normalized to 6 ( = the number of degrees of freedom per elementary cell) DOS is given by the formula
\begin{equation}\label{second}
\begin{array}{l}
\mathcal{R}(\lambda)= \frac{1}{\pi}\lim\limits_{\varepsilon\downarrow 0} \left(\frac{a}{2\pi}\right)^2  \int\limits_{-\pi/a}^{\pi/a}\int\limits_{-\pi/a}^{\pi/a} \sum\limits_{j=1}^{6} \frac{\varepsilon}{\left(\lambda -\omega_{j}^{2}(k_{1},k_{2})\right)^{2}+\varepsilon^{2}} dk_{1} dk_{2}= \\
\qquad\quad\; \frac{1}{\pi}\lim\limits_{\varepsilon\downarrow 0} \,\mathrm{Im}\left(\frac{a}{2\pi}\right)^2  \int\limits_{-\pi/a}^{\pi/a}\int\limits_{-\pi/a}^{\pi/a} \Tr\left[\mathbf{D}(\mathbf{k})-(\lambda +\rmi\varepsilon )\mathcal{I}\right]^{-1}dk_{1} dk_{2},\\
\qquad\qquad\qquad\qquad \left(\int\limits_0^{\infty}\mathcal{R}(\lambda)d\lambda=6\right),
\end{array}
\end{equation}
where $\mathcal{I}$ is the $6\times6$ unity matrix.

In order to obtain the phonon dispersion curves and DOS for ideal graphene on the base of expressions (\ref{eigen}) and (\ref{second}) concrete numerical values of the force constants $J_{1},J_{2}, J_{3}$ are necessary. We derived these constants by the least-squares method using experimental values of graphene phonon frequencies at the $\Gamma$, K and M points measured using the method of inelastic x-ray scattering. From now on we assume that
\begin{equation}\label{const}
\begin{array}{l}
J_{1}=3.79 \times 10^{-21} J/m^2, \\ J_{2}=6.89 \times 10^{-21} J/m^2, \\ J_{3}=2.36 \times 10^{-21} J/m^2.
\end{array}
\end{equation}

With the constants (\ref{const}) we get the values of phonon frequencies at the $\Gamma$, K and M points of Brillouin zone (\Tref{Table2}) and the sound velocities (which are in a good agreement with \cite{Maultzsch,TA1,TA2})
\[\begin{array}{l}
c_{LA}=18.4 \,km/sec,\\ c_{TA}=16.5\,km/sec,\\ c_{ZA}=\;\; 9.2\,km/sec.
\end{array}
\]

\begin{figure}[!hbp]
\center
\includegraphics[scale=0.78]{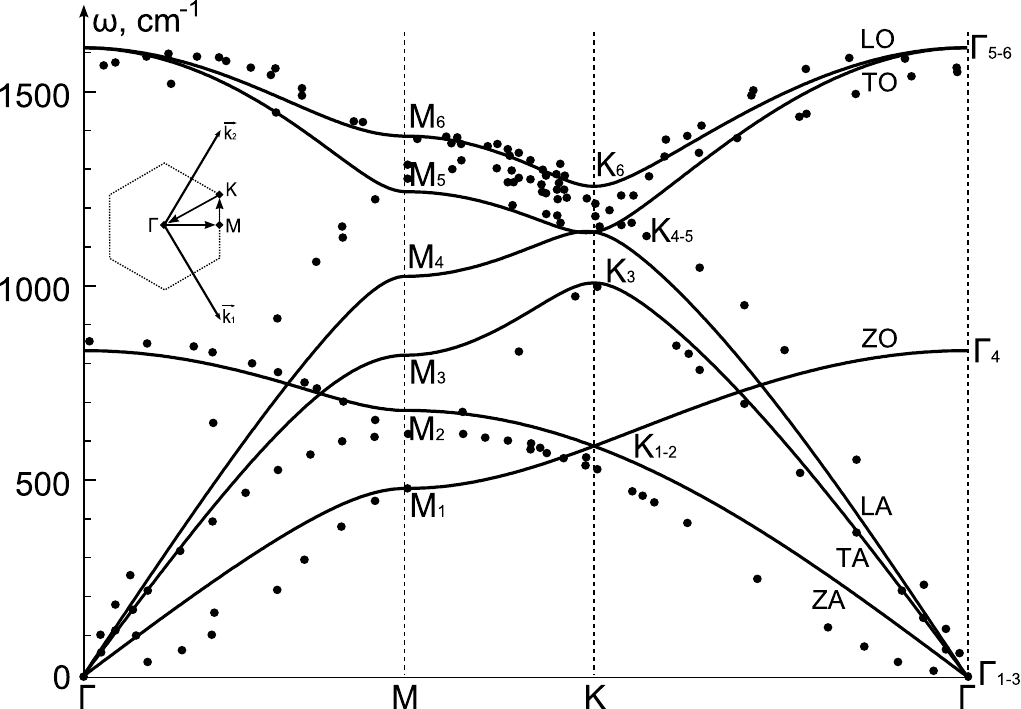}
\caption{Theoretical phonon frequency dispersion curves of the graphene along $\Gamma-\mathrm{M}-\mathrm{K}-\Gamma$ directions and experimental data for graphite (solid circles) \cite{Maultzsch}. LA (LO), TA (TO) and ZA (ZO) are longitudinal, transversal and out-of-plane acoustical (optical) branches respectively.} \label{picIdealDisp}
\end{figure}

\begin{figure}[!hbp]
\center
\includegraphics[scale=0.78]{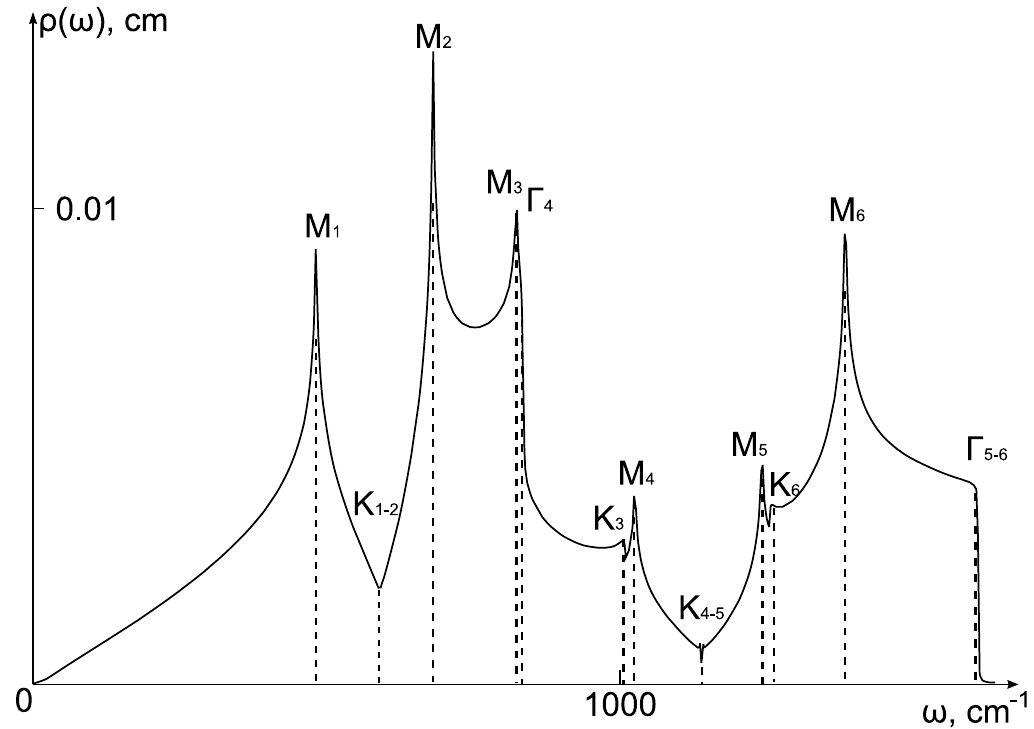}
\caption{The phonon density of states of the infinite graphene plane without defects. The phonon frequencies at the high-symmetry points of the Brillouin zone are labeled by $\Gamma_{i}$,M$\,_{i}$ and K$\,_{i}$.}
\label{picIdealDOS}
\end{figure}

For demonstration of how the fitted three-parameter model works we give below the phonon dispersion curves (figure \ref{picIdealDisp}) and density of states (figure \ref{picIdealDOS}) for ideal graphene.

\Table{\label{Table2} Theoretical phonon frequencies for the $\Gamma$, M and K points of the Brillouin zone.}
\br
\;& \centre{6}{Phonon frequencies ($cm^{-1}$)}  \\ \ns
\;& \centre{1}{$\omega_1$}& \centre{1}{$\omega_2$}& \centre{1}{$\omega_3$}& \centre{1}{$\omega_4$}& \centre{1}{$\omega_5$}& \centre{1}{$\omega_6$}\\
\mr
  $\Gamma$ &  0  &  0  &  0  & 840  & \centre{2}{1620} \\
  M        & 485 & 686 & 828 & 1031 & 1249 & 1392 \\
  K        & \centre{2}{594}  & 1014 & \centre{2}{1146} & 1263 \\
\br
\endTable

\section{Isotopic defects}

The kinetic energy $K$ and potential energy $W$ of the non-ideal graphene can be written as the following quadratic forms
\[\begin{array}{l} K=\frac{1}{2}\sum\limits_{\mathbf{l}}\sum\limits_{\sigma}^{}m(\mathbf{l})\mathbf{\dot{u}}_{\mathbf{l},\sigma}^{2} = \\
\qquad \frac{1}{2}\sum\limits_{\mathbf{l}}\sum\limits_{\sigma}^{}m(\mathbf{l}) \left(\dot{u}_{\mathbf{l}\sigma,1}^{2}+\dot{u}_{\mathbf{l}\sigma,2}^{2}+\dot{u}_{\mathbf{l}\sigma,1}\dot{u}_{\mathbf{l}\sigma,2}
+\dot{u}_{\mathbf{l}\sigma,3}^2 \right), \end{array} \]
\[
W\left(\{\mathbf{u}_{\sigma}(\mathbf{l})\}\right)=\frac{1}{2}\sum\limits_{\mathbf{l}',\mathbf{l}}^{}\sum\limits_{\sigma',\sigma=A}^{B}
\sum\limits_{j',j=1}^{3}c_{\mathbf{l}'\sigma' j',\mathbf{l}\sigma j}u_{\sigma' j'}(\mathbf{l}')u_{\sigma j}(\mathbf{l}),
\]
where $\mathbf{l}\sigma$ enumerate the graphene lattice sites. From now on we will denote by $\mathfrak{M}$  and $\mathfrak{C}$ the matrices of the quadratic forms $2K$ and $2W$, respectively. We will call $\mathfrak{M}$ the mass matrix and $\mathfrak{C}$ the force constant matrix.
For the case of the ideal graphene  $\mathfrak{M}$ is the block diagonal matrix $\mathfrak{M}_{0}$ with equal diagonal blocks
\begin{equation}\label{mass}
\mathbf{M}^{0}_{\mathbf{l}\sigma, \mathbf{l}\sigma}
=\mathbf{M}_{0}
\end{equation}
and $\mathfrak{C}$ is the matrix of the doubled sum of the quadratic forms (\ref{pot1}). For graphene with substitutional defects the mass matrix is block diagonal and all its diagonal blocks have form (\ref{mass}) but with masses of substitutional ions instead of $m_{0}$ for some $\mathbf{l}\sigma$.

Let $\mathfrak{M}$ be the mass matrix of a non-ideal graphene plane with some host $\,^{12}C$ atoms replaced by other stable carbon isotopes. Note that the block-diagonal matrices $\mathfrak{M}_{0}$ and $\mathfrak{M}$ commute. Let $\rho_{0}(\omega)$ and $\rho_{\mathfrak{M}}(\omega)$ denote DOS of carbon planes with mass matrices $\mathfrak{M}_{0}$ and $\mathfrak{M}$, respectively. As $\omega_{\nu}^{2}$ in (\ref{first}) are eigenvalues of the matrix \[\mathfrak{D}=\mathfrak{M}^{-\frac{1}{2}}\mathfrak{C}\,\mathfrak{M}^{-\frac{1}{2}}\]
or the linear pencil $\mathfrak{C}-z\mathfrak{M}$, i.e. $\omega_{\nu}^{2}$ are those values of $z$, for which $\mathfrak{C}-z\mathfrak{M}$ is non-invertible, then
\begin{equation}\label{second}
\begin{array}{l}
\mathcal{R}(\omega^{2})=\frac{1}{\pi}\lim\limits_{\varepsilon\downarrow 0}\,\mathrm{Im}\Tr\left[\mathfrak{D}-
\left(\omega^{2}+\rmi\varepsilon\right)\mathfrak{I}\right]^{-1}=\\
\quad\qquad\;\;\; \frac{1}{\pi}\lim\limits_{\varepsilon\downarrow 0}\, \mathrm{Im}\Tr[\mathfrak{C}-\left(\omega^{2}+\rmi\varepsilon\right)\mathfrak{M}]^{-1}\mathfrak{M},
\end{array}
\end{equation}
where $\mathfrak{I}$ is the unity matrix.

Note that for any positive definite matrices $\mathfrak{C}$ and $\mathfrak{M}$ of any finite order and non-real $z$ we can write
\begin{equation} \label{third}
\Tr\left(\left[\mathfrak{C}-z\mathfrak{M}\right]^{-1}\mathfrak{M}\right)=-\frac{\rmd}{\rmd z}\Tr\ln\left(\mathfrak{C}-z\mathfrak{M}\right)=
-\frac{\rmd}{\rmd z}\ln \det(\mathfrak{C}-z\mathfrak{M}).
\end{equation}

Therefore from (\ref{first}) and (\ref{third}) we get
\begin{equation}\label{fourth} \fl \qquad
\rho_{\mathfrak{M}}(\omega) = \rho_{0}(\omega)-\frac{2\omega}{\pi}\lim\limits_{\varepsilon\downarrow 0}\,\mathrm{Im} \left\{\frac{\rmd}{\rmd z} \ln\det\left[(\mathfrak{C}-z\mathfrak{M})(\mathfrak{C}-z\mathfrak{M}_{0})^{-1}\right]\right\}_{z=\omega^{2}+\rmi\varepsilon}
\end{equation}
or
\begin{equation}\label{fifth}
\begin{array}{l}
\rho_{\mathfrak{M}}(\omega)-\rho_{0}(\omega)=\\
\;\; -\frac{2\omega}{\pi}\lim\limits_{\varepsilon\downarrow 0}\mathrm{Im}\frac{\rmd}{\rmd z}\ln\det \left[\mathfrak{I}-z\left(\mathfrak{M}-\mathfrak{M}_{0}\right)\left(\mathfrak{C}-z\mathfrak{M}_{0}\right)^{-1}\right]|_{z=\omega^{2}+\rmi\varepsilon}=\\ \;\; -\frac{2\omega}{\pi}\lim\limits_{\varepsilon\downarrow 0}\mathrm{Im}\frac{\rmd}{\rmd z}\ln\det \left[\mathfrak{I}-z\left(\mathfrak{M}\mathfrak{M}_{0}^{-1}-\mathfrak{I}\right)
\left(\mathfrak{D}-z\mathfrak{I}\right)^{-1}\right]\mathbb{}|_{z=\omega^{2}+\rmi\varepsilon}.
\end{array}
\end{equation}

From now on we will assume, for simplicity, that only one species of isotopes with mass $m$ can replace the host atoms of the ideal graphene lattice and assign to each site $\mathbf{l}\sigma$ of non-ideal graphene the occupation number
\begin{equation}\label{seventh}
n_{\mathbf{l}\sigma}=\left\{ \begin{array}{l}
                                0 \quad \textrm{if the host carbon atom seats at}\,\, \mathbf{l}\sigma; \\
                                1 \quad \textrm{if the carbon isotope is there}.
                             \end{array} \right.
\end{equation}

We can express diagonal blocks of $\mathfrak{M}$ in terms of $n_{\mathbf{l}\sigma}$ as follows
\begin{equation}\label{mass1}
\mathbf{M}_{\mathbf{l}\sigma, \mathbf{l}\sigma}=\left[1+\frac{m-m_{0}}{m_{0}}n_{\mathbf{l}\sigma}\right] \mathbf{M}_{0}.
\end{equation}
By (\ref{mass1}) $\left(\mathfrak{M}\mathfrak{M}_{0}^{-1}-\mathfrak{I}\right)$ is a block diagonal matrix with $3\times3$ diagonal blocks of the form
\begin{equation}\label{mass2}
\mathbf{\Lambda}_{\mathbf{l}\sigma, \mathbf{l}\sigma}
=\mu \cdot n_{\mathbf{l}\sigma} \mathbf{I},  \qquad \mu=\frac{m-m_{0}}{m_{0}},
\end{equation}
where $\mathbf{I}$ is the $3\times3$ unity matrix.

In view of the obvious property of occupation numbers $n_{\mathbf{l}\sigma}^{2}=n_{\mathbf{l}\sigma}$ we can represent $\rho_{\mathbf{B}}(\omega)$ formally as follows:
\begin{equation}\label{decomp}
\begin{array}{l}
\rho_{\mathfrak{M}}(\omega)=\rho_{0}(\omega)+\frac{1}{1!}\sum\limits_{\mathbf{l}\sigma}^{}\Xi_{1}(\omega ;\mathbf{l}\sigma)n_{\mathbf{l}\sigma}+\frac{1}{2!}\sum\limits_{\mathbf{l}\sigma\neq\mathbf{l}'\sigma'}^{}\Xi_{2}(\omega ;\mathbf{l}\sigma,\mathbf{l}'\sigma')n_{\mathbf{l}\sigma}n_{\mathbf{l}'\sigma'}+\\
\qquad\qquad \frac{1}{3!}\sum\limits_{\mathbf{l}\sigma\neq\mathbf{l}'\sigma'\neq\mathbf{l}"\sigma"}^{}
\Xi_{3}(\omega;\mathbf{l}\sigma,\mathbf{l}'\sigma',\mathbf{l}''\sigma'')n_{\mathbf{l}\sigma}n_{\mathbf{l}'\sigma'}
n_{\mathbf{l}''\sigma''}+...   \end{array}
\end{equation}
In fact, the decomposition (\ref{decomp}) is an identity, which holds for any number of isotopes (or other point defects) and their distribution over the graphene lattice sites. Particularly, if there is only one isotopic defect located at the lattice site $\mathbf{l}_{0}\sigma_{0}$, that is if $n_{\mathbf{l}\sigma}=\delta_{\mathbf{l}\sigma,\mathbf{l}_{0}\sigma_{0}}$, then according to (\ref{decomp})
\begin{equation}\label{delta1}
\Xi_{1}(\omega ;\mathbf{l}_{0}\sigma_{0})=\rho(\omega;\mathbf{l}_{0}\sigma_{0})-\rho_{0}(\omega),
\end{equation}
where $\rho(\omega;\mathbf{l}_{0}\sigma_{0})$ is the DOS of graphene lattice with a single defect at the site $\mathbf{l}_{0}\sigma_{0}$. In much the same way we find that
\begin{equation}\label{delta2}
\Xi_{2}(\omega;\mathbf{l}\sigma,\mathbf{l}'\sigma')=\rho(\omega;\mathbf{l}\sigma,\mathbf{l}'\sigma')-\Xi_{1}(\omega ;\mathbf{l}\sigma)-\Xi_{1}(\omega ;\mathbf{l}'\sigma')-\rho_{0}(\omega),
\end{equation}
where $\rho(\omega;\mathbf{l}\sigma,\mathbf{l}'\sigma')$ is the DOS of graphene lattice with only two isotopic defects at the sites $\mathbf{l}\sigma$ and $\mathbf{l}'\sigma'$ and so on.

Let $\Gamma(z)$ denote the $3\times3$ diagonal block of $\left(\mathfrak{D}-z\mathfrak{I}\right)^{-1}$ with some index $\mathbf{l}\sigma$. By the translational and point symmetry of graphene plain $\Gamma(z)$ does not depend on $\mathbf{l}\sigma$. It follows from (\ref{fifth}) and (\ref{mass1}) that the coefficients $\Xi_{1}(\omega ;\mathbf{l}_{0}\sigma_{0})$ in (\ref{decomp}) actually do not depend on $\mathbf{l}$ and $\sigma$,
\[
\Xi_{1}(\omega)\left(=\Xi_{1}(\omega ;\mathbf{l}_{0}\sigma_{0})\right)=-\frac{2\omega}{\pi}\lim\limits_{\varepsilon\downarrow 0}\mathrm{Im} \frac{\rmd}{\rmd z}\ln\det\left[\mathbf{I}-\mu z\Gamma(z)\right] \mathbb{}|_{z=\omega^{2}+\rmi\varepsilon}.
\]

Note that for $3\times3$ blocks $\mathbf{d}_{\mathbf{l}\sigma,\mathbf{l}'\sigma'}$ of $\mathfrak{D}=\left(\mathfrak{M}_{0}^{-\frac{1}{2}}\mathfrak{C}\,\mathfrak{M}_{0}^{-\frac{1}{2}}\right)$ we have
\[\begin{array}{cc}
    \mathbf{d}_{\mathbf{l}\sigma;\mathbf{l}'\sigma'}=\mathbf{d}_{\mathbf{l}-\mathbf{l}'\sigma;\mathbf{0}\sigma'}; & \mathbf{d}_{\mathbf{l}-\mathbf{l}'\sigma;\mathbf{0}\sigma}=\mathbf{d}_{\mathbf{l}-\mathbf{l}'\tau;\mathbf{0}\tau}, \; \sigma\neq\tau.
  \end{array}
\]
Making use of the explicit expression for the dynamical matrix $\mathcal{\mathbf{D}}(\mathbf{k})=\left(\mathcal{D}_{\sigma\tau}(\mathbf{k})\right)_{\sigma,\tau=1}^{2}$ of the ideal graphene plane
\begin{equation}\label{dynamic}
\mathcal{D}_{\sigma\tau}(\mathbf{k})=\frac{1}{m_{0}}\sum\limits_{\mathbf{l}}^{}\mathbf{d}_{\mathbf{l}\sigma;\mathbf{0}\tau}%
\rme^{-\rmi\mathbf{k}\cdot\mathbf{l}},
\end{equation}
where $\mathbf{k}$ is the wave vector from the Brillouin zone, we obtain that
\begin{equation}\label{gamma3}
\Gamma(z)=\frac{1}{N}\sum\limits_{\mathbf{k}}^{}\left(\left[\mathcal{D}(\mathbf{k})-z\mathcal{I}\right]^{-1}\right)_{\sigma\sigma},
\end{equation}
where $N$ is the number of unit cells in the periodicity area. Hence
\begin{equation}\label{xi1} \fl
\Xi_{1}(\omega)=-\frac{2\omega}{\pi}\lim\limits_{\varepsilon\downarrow 0}\mathrm{Im}\left. \frac{\rmd}{\rmd z}\ln\det\left[\mathbf{I}-\mu z\frac{1}{N}\sum\limits_{\mathbf{k}}^{}\left(\left[\mathcal{D}(\mathbf{k})-z\mathcal{I}\right]^{-1}\right)_{\sigma\sigma}\right] \right|_{z=\omega^{2}+\rmi\varepsilon}.
\end{equation}

Proceeding in the same fashion and setting
 \begin{equation}\label{gamma4}
\Gamma_{\sigma\tau}(z;\mathbf{l})=\frac{1}{N}\sum\limits_{\mathbf{k}}^{}\rme^{\rmi\mathbf{k}\cdot\mathbf{l}} \left(\left[\mathcal{D}(\mathbf{k})-z\mathcal{I}\right]^{-1}\right)_{\sigma\tau}
\end{equation}
we find that
\begin{equation}\label{xi2} \fl
\begin{array}{l}
 \Xi_{2}(\omega;\mathbf{l}\sigma,\mathbf{l}'\tau)=\Xi_{2}(\omega;\mathbf{l}-\mathbf{l}'\sigma,\mathbf{0}\tau)= \\
  -2\Xi_{1}(\omega)-\frac{2\omega}{\pi}\lim\limits_{\varepsilon\downarrow 0}\mathrm{Im}\frac{\rmd}{\rmd z}\ln\det\left(
  \begin{array}{cc}
   \mathbf{I}-z\mu\Gamma(z) & -z\mu\Gamma_{\sigma\tau}(z;\mathbf{l}-\mathbf{l}') \\
   -z\mu\Gamma_{\tau\sigma}(z;\mathbf{l}'-\mathbf{l}) & \mathbf{I}-z\mu\Gamma(z)
 \end{array} \right)_{z=\omega^{2}+\rmi\varepsilon} .
\end{array}
\end{equation}

\begin{figure}[!hbp]
\center
\includegraphics[scale=0.95]{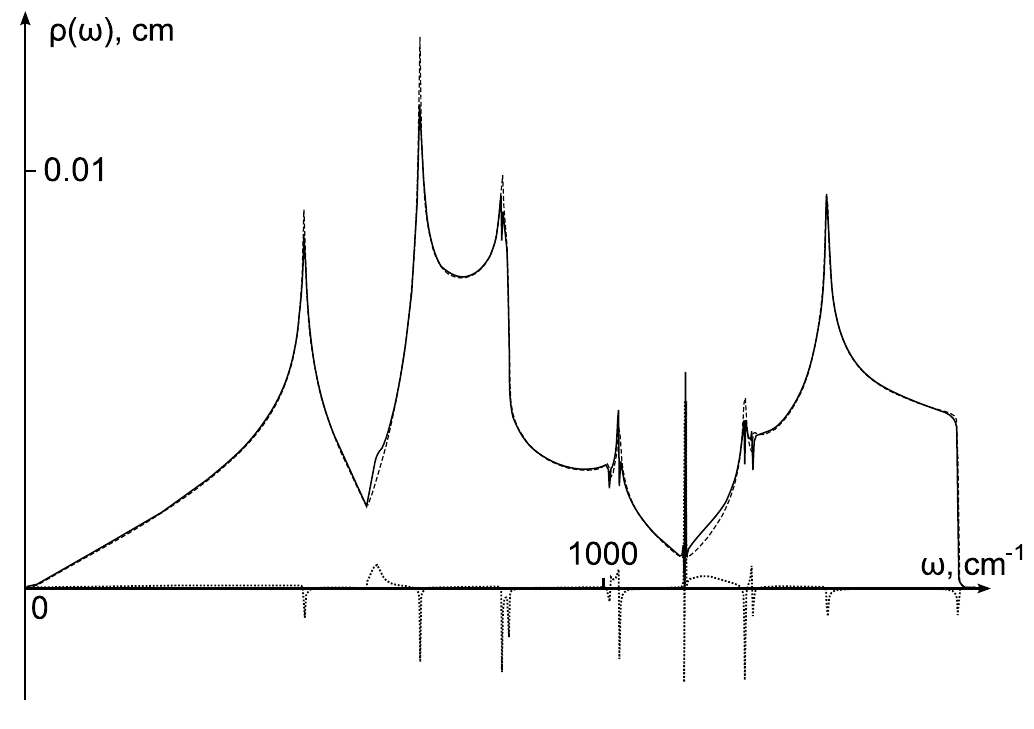}
\caption{The influence of 2\% aluminium defect concentration (isotopic case) on the ideal graphene phonon density of states. The dashed, solid and dotted lines represent DOSs of the ideal graphene, defect graphene (2\% of Al atoms) and defect contribution respectively. }\label{picGrapheneWithALisotopic}
\end{figure}

\begin{figure}[!hbp]
\center
\includegraphics[scale=0.95]{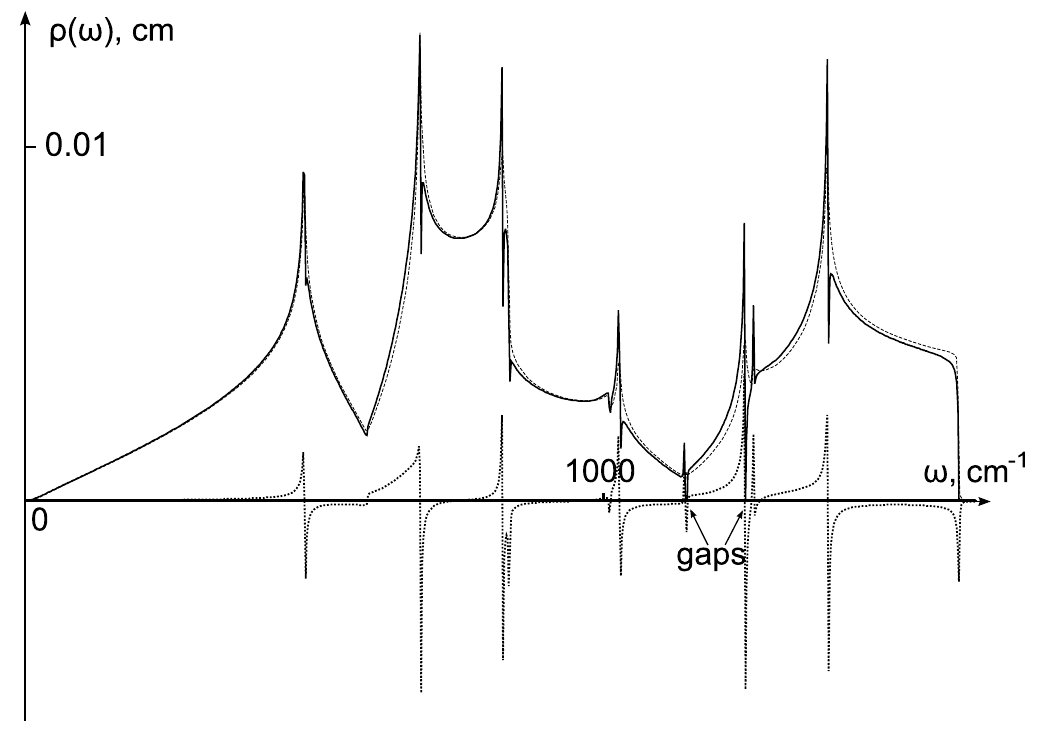}
\caption{The influence of 20\% nitrogen doping (in the case of dimers) on the ideal graphene phonon density of states. The dashed, solid and dotted lines represent DOSs of the ideal graphene, defect graphene (20\% of N atoms) and defect contribution respectively. Two gaps in the density of states near $\omega_{K_{4,5}}$ and $\omega_{M_{6}}$ are labeled by the arrows.}\label{picGrapheneWithNDimers}
\end{figure}

We pay now attention to the fact that for the graphene plane of finite size $\rho_{\mathfrak{M}}(\omega),\rho_{0}(\omega)$ in  (\ref{decomp}) have order $N$ while the coefficients $\Xi_{1}(\omega),\Xi_{2}(\omega;\mathbf{l}\sigma,\mathbf{0}\tau),...$ are finite quantities. Let us consider the equilibrium distribution of defects (isotopes) with concentrations {c} and the pair distribution function of defects
\[
g_{\sigma\tau}(\mathbf{l})=\lim\limits_{N\rightarrow\infty}N\langle n_{\mathbf{l}\sigma}n_{\mathbf{0}\tau}\rangle,
\]
where angle brackets denote thermal (or other) average. Then for the phonon spectral densities per unit cell of the graphene plane
\[
\Delta_{\mathfrak{K}}(\omega)=\lim\limits_{N\rightarrow\infty}\frac{1}{N}\langle \rho_{\mathfrak{K}}(\omega)\rangle, \:  \mathfrak{K}=\mathfrak{M},0 ,
\]
by (\ref{decomp}), (\ref{xi1}), (\ref{xi2}) we have
\begin{equation}\label{average}
\Delta_{\mathfrak{M}}(\omega)=\Delta_{0}(\omega)+c\Xi_{1}(\omega)+\frac{1}{2!}c^{2}\sum\limits_{\mathbf{l}\sigma\tau}^{}\Xi_{2}(\omega;\mathbf{l}\sigma,\mathbf{0}\tau)%
g_{\sigma\tau}(\mathbf{l})+...
\end{equation}
If for some reason there is a visible concentration $p$ of isotopic dimers on the graphene plane, that is cells occupied by pair of isotopes, then their contribution to $\Delta_{\mathfrak{M}}(\omega)$ is (figure \ref{picGrapheneWithNDimers})
\begin{equation}\label{dimer} \fl
 \begin{array}{l}
  \Delta_{\mathfrak{M}}(\omega)=...+p\cdot\Xi_{\mathrm{dim}}(\omega)..., \\
  \Xi_{\mathrm{dim}}(\omega) = -\frac{2\omega}{\pi}\lim\limits_{\varepsilon\downarrow 0}\mathrm{Im}\frac{\rmd}{\rmd z}\ln\det\left(\!
  \begin{array}{cc}
   \mathbf{I}-z\mu\Gamma(z) & -z\mu\Gamma_{\sigma\tau}(z;\mathbf{0}) \\
   -z\mu\Gamma_{\tau\sigma}(z;\mathbf{0}) & \mathbf{I}-z\mu\Gamma(z)
  \end{array}\! \right)_{z=\omega^{2}+\rmi\varepsilon}.
 \end{array}
\end{equation}

\section{Substitutional atoms}
Let us consider now a non-ideal graphene plane with low concentration $c_{X}$ of impurity atoms $X$ replacing carbon atoms at some lattice sites. Applying the same arguments as above we can again assert that in this case the phonon spectral density per unit cell $\Delta_{X}(\omega)$ can be written as follows:
\begin{equation}\label{averageX}
\Delta_{X}(\omega)=\Delta_{0}(\omega)+c_{X}\Theta_{1}(\omega)+\frac{1}{2!}c_{X}^{2}\sum\limits_{\mathbf{l}\sigma\tau}^{}\Theta_{2}(\omega;\mathbf{l}\sigma,\mathbf{0}\tau)%
g^{X}_{\sigma\tau}(\mathbf{l})+o(c_{X}^{2}),
\end{equation}
Here
\begin{equation}\label{delta3}
\Theta_{1}(\omega)=\rho_{X}(\omega)-\rho_{0}(\omega),
\end{equation}
$\rho_{X}(\omega)$ being the DOS of graphene lattice with a single defect at some site $\mathbf{l}_{0}\sigma_{0}$; \begin{equation}\label{delta4}
\Theta_{2}(\omega;\mathbf{l}\sigma,\mathbf{0}\sigma')=\rho_{X}(\omega;\mathbf{l}\sigma,\mathbf{0}\sigma')-2\Theta_{1}(\omega) +\rho_{0}(\omega),
\end{equation}
$\rho_{X}(\omega;\mathbf{l}\sigma,\mathbf{0}\sigma')$ being the DOS of graphene lattice with only two impurity atoms $X$ at some sites $\mathbf{l}'\sigma,\,\mathbf{l}''\sigma'$ such that $\mathbf{l}=\mathbf{l}'-\mathbf{l}''$.
Denote by $\mathfrak{M}_{X}$ the mass matrix and by $\mathfrak{C}_{X}$ the force constants matrix of the graphene lattice with a small number of impurity atoms $X$ replacing carbon at some sites and put
\[
\begin{array}{cc}
 \delta\mathfrak{M}_{X}=\mathfrak{M}_{X}- \mathfrak{M}_{0},& \delta\mathfrak{C}_{X}=\mathfrak{C}_{X}- \mathfrak{C}.
\end{array}
\]

As a starting point for calculation of $\Theta_{1}(\omega), \Theta_{2}(\omega;\mathbf{l}\sigma,\mathbf{0}\sigma'),...$ we make use of the relation
\begin{equation}\label{newfifth}
\begin{array}{l}
\rho_{X}(\omega)-\rho_{0}(\omega)=\\
\;\; -\frac{2\omega}{\pi}\lim\limits_{\varepsilon\downarrow 0}\mathrm{Im}\frac{\rmd}{\rmd z} \ln\det\left[\mathfrak{I}+\left(\delta\mathfrak{C}_{X}-z\delta\mathfrak{M}_{X}\right)
\left(\mathfrak{C}-z\mathfrak{M}_{0}\right)^{-1}\right]|_{z=\omega^{2}+\rmi\varepsilon}.
\end{array}
\end{equation}
Note that as well as for the graphene lattice with only isotopic defects the matrix
\[
T_{X}(z):=\mathfrak{M}^{-\frac{1}{2}}\left( \delta\mathfrak{C}_{X}-z\delta\mathfrak{M}_{X} \right) \mathfrak{M}^{-\frac{1}{2}}
\]
can be represented in the block form, elements of which are $3\times3$ matrices being enumerated by pairs of the multi-indices $\mathbf{l}\sigma,\,\mathbf{l}'\tau$ that enumerate the graphene lattice sites. Actually, only those blocks of $T_{X}(z)$ are non-zero, for which each of the indices $\mathbf{l}\sigma,\,\mathbf{l}'\tau$ belongs to the subset of indices $\mathfrak{D}$ of those enumerating either sites occupied by $X$-atoms or their nearest neighbors. Therefore if there are $N_{X}$ impurities in the periodicity area, then the rank of $T_{X}(z)$ does not exceed $12N_{X}$. For these reasons the determinant in (\ref{newfifth}) can be replaced by the minor determinant obtained from $\det(\mathfrak{I}+...)$ by deleting all rows and columns with "numbers" other then those from $\mathfrak{D}$. By this argument setting
\[
\begin{array}{cc}
\mathcal{Q}(z)=\left[\Gamma_{\sigma\tau}(z;\mathbf{l}-\mathbf{l}')\right]_{\mathbf{l}\sigma,\mathbf{l}'\tau \in
\mathfrak{D}},& \mathcal{T}_{X}(z)=\left[ T_{X}(z)_{\mathbf{l}\sigma,\mathbf{l}'\tau}\right]_{\mathbf{l}\sigma,\mathbf{l}'\tau\in\mathfrak{D}}
\end{array}
\]
we can write
\begin{equation}\label{newfifth1}
\begin{array}{l}
\rho_{X}(\omega)-\rho_{0}(\omega)=\\
\;\; -\frac{2\omega}{\pi}\lim\limits_{\varepsilon\downarrow 0}\mathrm{Im}\left.\frac{\rmd}{\rmd z}
\ln\det\left\{\mathcal{T}_{X}(z)\left[\mathcal{T}_{X}(z)^{-1}+\mathcal{Q}(z)\right]^{-1}\right\}\right|_{z=\omega^{2}+\rmi\varepsilon}.
\end{array}
\end{equation}

Let us assume for certainty that the single defect is located at the site $\mathbf{0}A$ and therefore its "normal" nearest neighbors are located at the sites $\mathbf{l}_{1}B$, $\mathbf{l}_{2}B$, $\mathbf{l}_{3}B$, where $\mathbf{l}_{1}=\mathbf{0}$, $\mathbf{l}_{2}=-\mathbf{a}_{1}$, $\mathbf{l}_{3}=-\mathbf{a}_{2}$, respectively. In this case the subset introduced above
\[
\mathfrak{D}=\left\{\mathbf{0}A, \; \mathbf{l}_{1}B, \; \mathbf{l}_{2}B, \; \mathbf{l}_{3}B, \; \mathbf{l}_{1}B\right\}.
\]
We may count further for brevity $\mathbf{0}$ instead of $\mathbf{0}A$ and $\mathbf{s}$ instead of $\mathbf{l}_{s}B$, $s=1,2,3$.
\begin{figure}[!hbp]
\center
\includegraphics[scale=0.90]{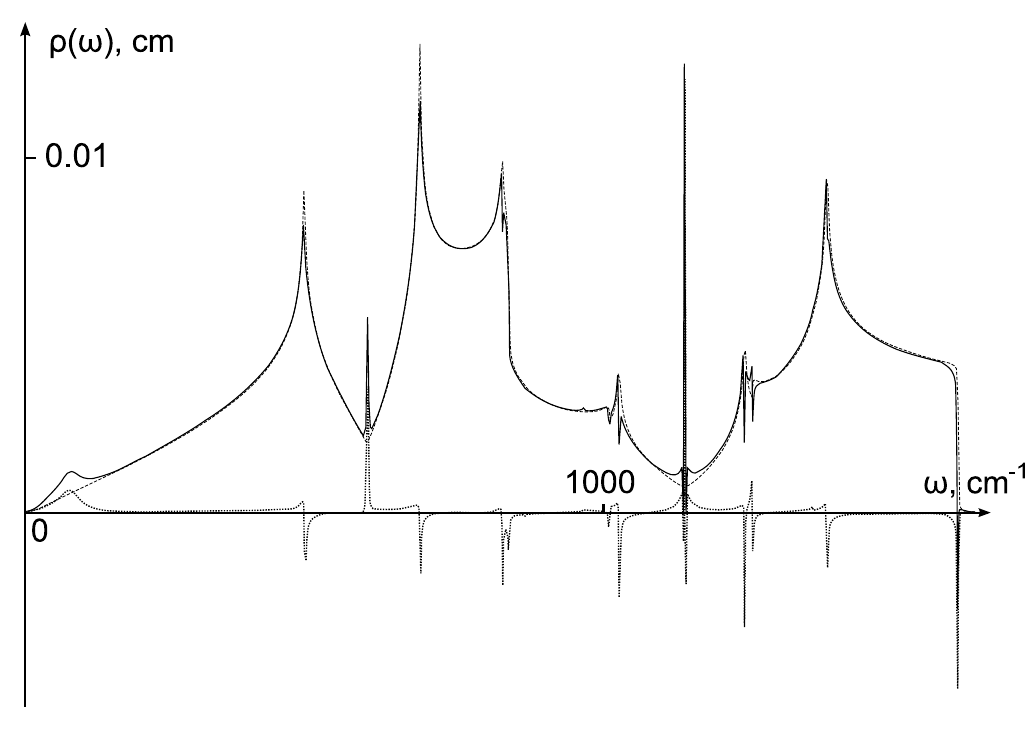}
\caption{The influence of 2\% aluminium defect concentration (isobaric case, $J_{i}=0.5 J_{i,0}$) on the ideal graphene phonon density of states. The dashed, solid and dotted lines represent DOSs of the ideal graphene, defect graphene (2\% of Al atoms) and defect contribution respectively. }\label{picGrapheneWithALisobaric}
\end{figure}

\begin{figure}[!hbp]
\center
\includegraphics[scale=0.90]{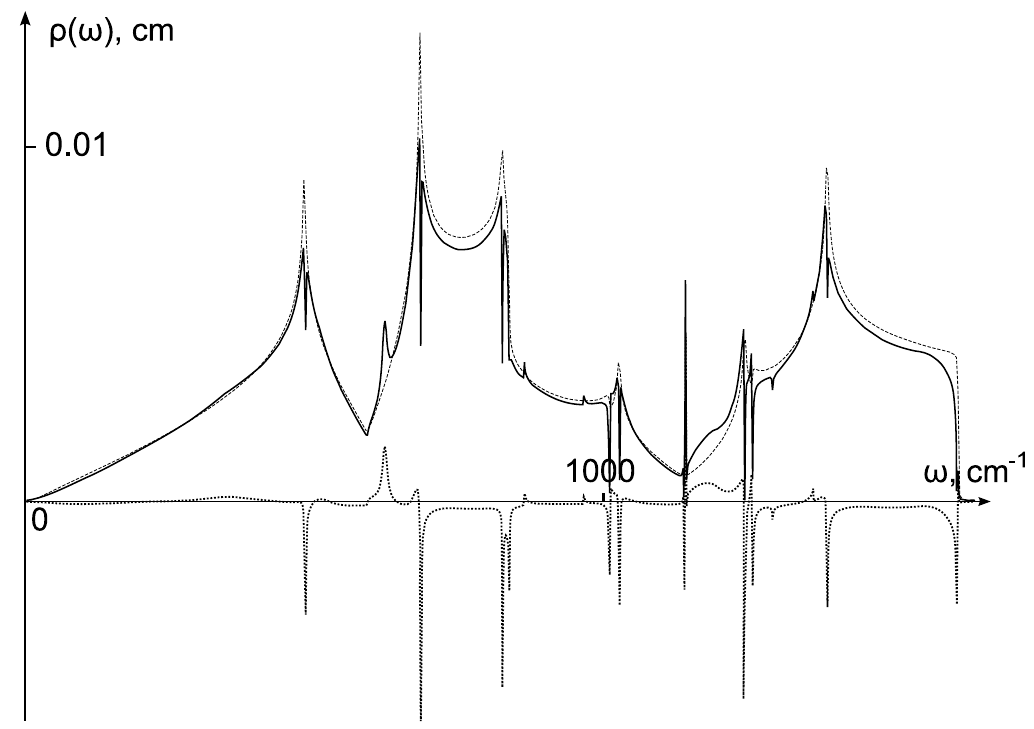}
\caption{The influence of 2.5\% of vacancies on the ideal graphene phonon density of states. The dashed, solid and dotted lines represent DOSs of the ideal graphene, defect graphene (2.5\% of vacancies) and defect contribution respectively.}\label{picGrapheneWithVacancies}
\end{figure}

With this enumeration the $12\times12$ matrices $\mathcal{T}_{X}(z)$ and $\mathcal{T}_{X}(z)^{-1}$ can be now represented as $4\times4$ block matrices
\begin{equation}\label{def2}
\begin{array}{l}
  \mathcal{T}_{X}(z)=\left(
                  \begin{array}{cccc}
                    -z\delta \mathbf{G}+\sum\limits_{s=1}^{3}\delta \mathbf{G}_{s} & - \delta \mathbf{G}_{1}& \delta \mathbf{G}_{2}& \delta \mathbf{G}_{3}\\
                    -\delta \mathbf{G}_{1} & \delta \mathbf{G}_{1} & 0 & 0 \\
                   - \delta \mathbf{G}_{2} & 0 & \delta \mathbf{G}_{2} & 0 \\
                    -\delta \mathbf{G}_{3}& 0 & 0 & \delta \mathbf{G}_{3} \\
                  \end{array}
                \right), \\
                \mathcal{T}_{X}(z)^{-1}=-\frac{1}{z \mu}\left(
                                              \begin{array}{cccc}
                                               \mathbf{I} & \mathbf{I} & \mathbf{I} & \mathbf{I} \\
                                               \mathbf{I} & \mathbf{I} & \mathbf{I} & \mathbf{I} \\
                                               \mathbf{I} & \mathbf{I} & \mathbf{I} & \mathbf{I} \\
                                               \mathbf{I} & \mathbf{I} & \mathbf{I} & \mathbf{I} \\
                                              \end{array}
                                         \right)+\left(
                                              \begin{array}{cccc}
                                               0 & 0 & 0 & 0 \\
                                               0 & \delta \mathbf{G}_{1}^{-1} & 0 & 0 \\
                                               0 & 0 & \delta \mathbf{G}_{2}^{-1} & 0 \\
                                               0 & 0 & 0 & \delta \mathbf{G}_{3}^{-1}\\
                                              \end{array} \right),
\end{array}
\end{equation}
where
\begin{equation}\label{mass3}
\delta \mathbf{M} =\mu \mathbf{I}
\end{equation}
and
\begin{equation}\label{force3}\fl
\begin{array}{l}
  \delta \mathbf{G}_{1}=\frac{1}{2} \frac{\Delta J_{1}}{m_{0}}\left(
                                        \begin{array}{ccc}
                                          1 & 1 & 0 \\
                                          1 & 1 & 0 \\
                                          0 & 0 & 0 \\
                                        \end{array}
                                      \right)+\frac{\Delta J_{2}}{m_{0}}\left(
                                                \begin{array}{ccc}
                                                  1 & 0 & 0\\
                                                  0 & 1 & 0 \\
                                                  0 & 0 & 0 \\
                                                \end{array}
                                              \right)+\frac{\Delta J_{3}}{m_{0}}\left(
                                                                    \begin{array}{ccc}
                                                                      0 & 0 & 0 \\
                                                                      0 & 0 & 0 \\
                                                                      0 & 0 & 1 \\
                                                                    \end{array}
                                                                  \right),
   \\
   \delta \mathbf{G}_{2}=\frac{1}{4}\frac{\Delta J_{1}}{m_{0}}\left(
                                        \begin{array}{ccc}
                                          2 + \!\sqrt{3} & -1 & 0 \\
                                          -1 & 2 - \!\sqrt{3} & 0 \\
                                          0 & 0 & 0 \\
                                        \end{array}
                                      \right)+\frac{\Delta J_{2}}{m_{0}}\left(
                                                \begin{array}{ccc}
                                                  1 & 0 & 0\\
                                                  0 & 1 & 0 \\
                                                  0 & 0 & 0 \\
                                                \end{array}
                                              \right)+\frac{\Delta J_{3}}{m_{0}}\left(
                                                                    \begin{array}{ccc}
                                                                      0 & 0 & 0 \\
                                                                      0 & 0 & 0 \\
                                                                      0 & 0 & 1 \\
                                                                    \end{array}
                                                                  \right),
\\
   \delta \mathbf{G}_{3}=\frac{1}{4}\frac{\Delta J_{1}}{m_{0}}\left(
                                        \begin{array}{ccc}
                                          2 - \!\sqrt{3} & -1 & 0 \\
                                          -1 & 2 + \!\sqrt{3} & 0 \\
                                          0 & 0 & 0 \\
                                        \end{array}
                                      \right)+\frac{\Delta J_{2}}{m_{0}}\left(
                                                \begin{array}{ccc}
                                                  1 & 0 & 0 \\
                                                  0 & 1 & 0 \\
                                                  0 & 0 & 0 \\
                                                \end{array}
                                              \right)+\frac{\Delta J_{3}}{m_{0}}\left(
                                                                    \begin{array}{ccc}
                                                                      0 & 0 & 0 \\
                                                                      0 & 0 & 0 \\
                                                                      0 & 0 & 1 \\
                                                                    \end{array}\right)
\end{array}
\end{equation}
By (\ref{def2}) and (\ref{mass3})  we have
\begin{equation}\label{det2}
\det\mathcal{T}_{X}(z)=-\left(\det\prod\limits_{s=1}^{3}\delta G_{s}\right)\mu^{3}z^{3}.
\end{equation}
Substitution of the above given expressions (\ref{def2}) - (\ref{force3}) and
\begin{equation}\label{Green}\fl \quad
 \mathcal{Q}(z)=\left(
                  \begin{array}{cccc}
                    \Gamma_{AA}(z;\mathbf{0}) & \Gamma_{AB}(z;\mathbf{0}) & \Gamma_{AB}(z;\mathbf{a}_{1}) & \Gamma_{AB}(z;\mathbf{a}_{2}) \\
                    \Gamma_{BA}(z;\mathbf{0}) & \Gamma_{BB}(z;\mathbf{0}) & \Gamma_{BB}(z;\mathbf{a}_{1}) & \Gamma_{BB}(z;\mathbf{a}_{2}) \\
                    \Gamma_{BA}(z;-\mathbf{a}_{1}) & \Gamma_{BB}(z;-\mathbf{a}_{1}) & \Gamma_{BB}(z;\mathbf{0}) & \Gamma_{BB}(z;\mathbf{a}_{2}\!-\! \mathbf{a}_{1})\\
                    \Gamma_{BA}(z;-\mathbf{a}_{2}) & \Gamma_{BB}(z;-\mathbf{a}_{2}) & \Gamma_{BB}(z;\mathbf{a}_{1}\!-\!\mathbf{a}_{2}) & \Gamma_{BB}(z;\mathbf{0}) \\
                  \end{array}
                \right)
\end{equation}
into (\ref{newfifth1}) gives the sought expression for the "coefficient" $\Theta_{1}(\omega)$ in (\ref{averageX}).

In the partial case of "isobaric" defect ($\mu=0$) we have
\begin{equation} \label{isobar}
 \begin{array}{l}
   \Theta_{1}(\omega)=-\frac{2\omega}{\pi}\lim\limits_{\varepsilon\downarrow 0}\mathrm{Im}\frac{\rmd}{\rmd z}
\ln\det\left(\mathfrak{I}+\mathcal{T}_{0}\mathcal{Q}(z)\right)|_{z=\omega^{2}+\rmi\varepsilon}, \\
 \mathcal{T}_{0} =
  \left(
   \begin{array}{cccc}
   \sum\limits_{s=1}^{3}\delta \mathbf{G}_{s} & - \delta \mathbf{G}_{1}& \delta \mathbf{G}_{2}& \delta \mathbf{G}_{3}\\
   -\delta \mathbf{G}_{1} & \delta \mathbf{G}_{1} & 0 & 0 \\
   -\delta \mathbf{G}_{2} & 0 & \delta \mathbf{G}_{2} & 0 \\
   -\delta \mathbf{G}_{3}& 0 & 0 & \delta \mathbf{G}_{3} \\
   \end{array}
  \right).
 \end{array}
\end{equation}

Not much remains to add to prove that for the vacancies at the graphene lattice sites the corresponding coefficient $\Theta_{1}^{0}(\omega)$ is given by the expression
\begin{equation}\fl\qquad
\begin{array}{l}
  \Theta_{1}^{0}(\omega)=-\frac{2\omega}{\pi}\lim\limits_{\varepsilon\downarrow 0}\mathrm{Im}\frac{\rmd}{\rmd z}
  \ln\det\mathcal{P}(z)|_{z=\omega^{2}+\rmi\varepsilon} \, , \\ \mathcal{P}(z)= \left(
        \begin{array}{ccc}
          \Gamma_{BB}(z;\mathbf{0}) - \delta \mathbf{G}_{1}^{0}& \Gamma_{BB}(z;\mathbf{a}_{1})  & \Gamma_{BB}(z;\mathbf{a}_{2})\\
          \Gamma_{BB}(z;-\mathbf{a}_{1}) & \Gamma_{BB}(z;\mathbf{0}) - \delta \mathbf{G}_{2}^{0}& \Gamma_{BB}(z;\mathbf{a}_{2}\!-\!\mathbf{a}_{1})\\
          \Gamma_{BB}(z;-\mathbf{a}_{2}) & \Gamma_{BB}(z;\mathbf{a}_{1}\!-\!\mathbf{a}_{2})  & \Gamma_{BB}(z;\mathbf{0}) - \delta \mathbf{G}_{3}^{0}\\
        \end{array}
  \right)\!,
\end{array}
\end{equation}
where $\delta \mathbf{G}_{1}^{0},\delta \mathbf{G}_{2}^{0},\delta \mathbf{G}_{3}^{0}$ are the particular cases of expressions (\ref{force3}) with $\Delta J_{1},\Delta J_{2},\Delta J_{3}$ replaced by $-J_{1},- J_{2},-J_{3}$, respectively.

\section{Discussion}
The proposed model for the description of phonon spectra in graphene, where the only  nearest neighbors interaction is accounted for, gives rather simple explicit expressions for the phonon dispersion curves and reflects quite satisfactory their main features for graphene and graphite in the whole frequency range. Certainly, containing only three force constant such a model can't reproduce all the specific features of graphite phonon spectra. For example, it cannot in principle reproduce the low-frequency bending mode with $\omega \sim k^2$ and the overbending of in-plane optical modes near the $\Gamma$ point so the more that the latter is assumed to be caused by the electron-phonon interaction \cite{CastroNeto,MaultzschEF,PiscanecEF}.

Nevertheless, the three force constants of the model being chosen as fitting parameters to obtain a good coincidence with known values (inelastic x-ray scattering on graphite \cite{Maultzsch}) of optical modes at the symmetrical points of the first Brillouin zone $\Gamma_{5,6}, \mathrm{K}_{6},\mathrm{K}_{4,5},\mathrm{K}_{3},\mathrm{M}_{6},\mathrm{M}_{5}$ (for in-plane modes) and $\Gamma_{4}, \mathrm{M}_{1}$ (for out-of-plane modes) give satisfactory (and sometimes rather good) quantitative agreement with experimental data for graphite throughout the whole frequency range. In particular, the value of in-plane transverse sound velocity agrees closely with the experimental value and results of the more sophisticated theoretical models \cite{Maultzsch,TA1,TA2}. We took slightly overstated in-plane $\Gamma$-point frequency $\omega_{\Gamma_{5,6}}=1620 cm^{-1}$ (instead of a value in between 1580 - 1595 $cm^{-1}$ according to known experiments) for better fitting of experimental data in a larger part of Brillouin zone and for a formal account of the in-plane optical modes overbending near the $\Gamma$-point.
Since the experimental data for K and M points of graphene phonon spectrum are rather poor and the selection rules are strictly obeyed due to the high crystalline quality of graphene we take the graphite phonon spectrum as a target of the fitting procedure, more so that the phonon dispersion curves for graphite and graphene are similar in a large part of the Brillouin zone \cite{HREELS}.

Another visible discrepancy between our theory and experimental data (in addition to those mentioned above) is the sufficiently lower in frequency in-plane longitudinal acoustic (LA) branch (figure \ref{picIdealDisp}) in the $\Gamma-\mathrm{M}$ direction and as a result its incorrect trend between M and K points. The analytically obtained frequencies $\omega_{K_6}$ and $\omega_{M_4}$ are connected in the model by the relation $\omega_{K_6}=\sqrt{\frac{3}{2}}\, \omega_{M_4}$ which is different from the x-ray experimental result $\omega_{K_6} \approx \omega_{M_4}$ \cite{Maultzsch}. However, this relation is in a better agreement with Raman experiments which give $\omega_{K_6} \approx \sqrt{1.6}\omega_{M_4}$ \cite{MaultzschD}. Also for the $\omega_{\Gamma_{5,6}}$ and $\omega_{K_{4,5}}$ we get  $\omega_{\Gamma_{5,6}}=\sqrt{2}\,\omega_{K_{4,5}}$, while x-ray and Raman experiments give $1.32$ and $1.25$, respectively, (instead of $1.41$ (or $1.38$ with account of our $\omega_{\Gamma_{5,6}}$ overestimation)). The next interesting observation is that the $\omega_{M_{5}},\omega_{K_{4,5}},\omega_{K_{6}}, \omega_{\Gamma_{5,6}}$ ratios to $\omega_{M_{4}}$ which are drastically different from the x-ray experiment results are in a good agreement with Raman experiments despite of essential difference in absolute values of some of those frequencies \cite{MaultzschD}. It is worth mentioning that the similar relations and ratios were obtained within a more detailed model with up-to-third nearest neighbors interactions accounted for \cite{Falkovsky}.

Note that since the most substantial spectrum changes due to defects were expected near the Van Hove singularities, all of which are reproduced at the proper places by our simple model, it is evident that further improvements and refinements were not dictated by the objectives and needs of this work.

The study of the influence of point defects on the graphene phonon spectra on the basis of our simplified model for the ideal graphene lattice oscillations showed, as might be expected, that the isotopic defects slightly downshift Van Hove singularities in the phonon DOS for heavier than carbon atom defects and upshift them for light defects. For the defects with mass of 11 u.amu (like $\,^{11}B$) the additional Van Hove peak appears on the upper edge of the ideal graphene phonon spectra. For lighter defects this peak splits out and corresponds to the localized oscillation mode with frequency above $\omega_{\Gamma_{5,6}}$. Such peaks may be related to the so called $D^{\prime}$ ($\omega=1620\,cm^{-1}$) and $2D^{\prime}$ ($\omega=3250\,cm^{-1}$) bands of the graphite Raman spectra \cite{DPrime1,FerraryRaman} and were directly observed for 2.66\% boron doped graphene \cite{DefectsBoron3p}.

Substitutional defects, for which the force constants for defect-carbon bonds are weaker than that for carbon-carbon bonds, may drastically change the phonon DOS in the low-frequency region (from 60 to 200 $cm^{-1}$) depending on the defect atom mass and weakened force constants. Note that a trace of an oscillation mode with a frequency of about $100\!-120\!\,\,cm^{-1}$ near the $\Gamma$-point was observed in graphite \cite{Maultzsch,HREELS,Gsquare}. It turns out in some cases that a defect concentration even of several per cent may result in 100\% low-frequency DOS increase in intervals wider than 100 $cm^{-1}$. Obviously, the changes of the phonon low-frequency DOS stipulated by defects should be manifested in the low temperature specific heat of non-ideal graphene. The isotopic defects with greater then $\,^{12}C$ mass lead to increase of the specific heat in proportion to the defect concentration (and temperature). Light defects slightly decrease the DOS and low temperature specific heat, but the difference from ideal graphene is sufficiently less than in the case of heavy defects. For substitutional defects the di®erence in density of states in the low-frequency region between graphene with defects and ideal graphene is much higher then in the case of isotopic defects and leads to significant changes in the specific heat (of order of $10\%$). The specific heat of 3\% Al doped graphene, $\Delta J_i/{J_i}=-0.25$, differs by more than 10\% of that of ideal graphene.

Apart from slight shifts of Van Hove points, significant changes in the DOSs are observed for any types of point defects near $\omega_{K_{1,2}}$ and $\omega_{K_{4,5}}$ points where oscillation modes are double-degenerated. For heavier isotopic defects the DOSs have more sharp peaks at the mentioned points. For substitutional atoms the changes near the $\omega_{K_{1,2}}$ and $\omega_{K_{4,5}}$ are similar to those for purely isotopic defects (figure \ref{picGrapheneWithALisobaric}). For example, the deviation from the ideal DOS can reach as high as 30\% for 2\% Al doped graphene (figure \ref{picGrapheneWithALisotopic}). In comparison with isotopic defects the same concentrations of vacant lattice sites lead to a greater DOS rise for frequencies from $\omega_{K_{4,5}}$ to $\omega_{M_{5}}$ and to sufficient DOS decrease in the intervals $(\omega_{M_{2}},\omega_{\Gamma_{4}})$ and $(\omega_{M_{6}},\omega_{\Gamma_{5,6}})$. That lowering may be of the order of several per cent for the wide frequency range (figure \ref{picGrapheneWithVacancies}). Both types of defect also lead to the high additional peaks in the $\omega_{K_{1,2}}$ (see also \cite{ltp2}) and $\omega_{K_{4,5}}$ (figures \ref{picGrapheneWithALisotopic}, \ref{picGrapheneWithALisobaric}). The singularities near the $\omega_{K_{1,2}}$ and $\omega_{K_{4,5}}$ may indicate the splitting of corresponding modes in the vicinity of the K-point and in the case of high defect concentration the appearance of gaps in the density of states near $\omega_{K_{4,5}}$ and $\omega_{M_{6}}$ is evident (figure \ref{picGrapheneWithNDimers}).

The vacant lattice sites except for the singularity at the $\omega_{K_{4,5}}$ point cause an additional resonant states peak with frequency between $\omega_{K_{1,2}}$ and $\omega_{M_{2}}$ (figure \ref{picGrapheneWithVacancies}).

In the ideal graphene the momentum conservation allows only one single-phonon Raman process which corresponds to the emission of the optical phonon with zero wave vector and frequency near $1580 cm^{-1}$ ($\omega_{\Gamma_{5,6}}$) \cite{Basko}. For non-ideal graphene there is also the so-called double-resonant presumably defect-induced peak (the D band) at $1350 cm^{-1}$ corresponding to emission of an optical phonons with wave vector near K points of the Brillouin zone \cite{Basko,DBand1,DBand2}. Within the commonly accepted interpretation, impurities only assist the photons scattering on single intervalley phonons. On the other hand, our analysis shows that all considered types of point defect cause additional Van Hove singularities of the DOS in the neighborhoods of the K and M points of optical branches, and that local oscillation modes may induce additional peaks of the Raman spectra.

% Acknowledgements
\ack
This work was supported by the Ministry of Education and Science of Ukraine, Grant \#0109U000929.

% ----------------------------------------------------------------------------------------------------------------------------------------------------

\end{document}